\documentclass[12pt]{article}

\usepackage{a4}
\usepackage{graphicx}
\usepackage{color}

\def\SU{\mathop{\rm SU}}
\def\U{\mathop{\rm U}}
\newcommand{\wt}{\widetilde}
\newcommand{\wh}{\widehat}
\newcommand{\ol}{\overline}

\def\tr{\mathop{\rm tr}\nolimits}

\newcommand{\cbox}[1]{\colorbox[rgb]{0.8,0.8,1}{#1}}

\def\citen{\cite}

\begin{document}

\begin{titlepage}
\title{\hfill\parbox{4cm}
       {\normalsize UT-04-27\\{\tt hep-th/0410138}\\October 2004}\\
       \vspace{1cm}
       On String Junctions\\in Supersymmetric Gauge Theories%
       \vspace{1cm}}
\author{Yosuke Imamura\thanks{E-mail: \tt imamura@hep-th.phys.s.u-tokyo.ac.jp}%
\\[20pt]
{\it Department of Physics, University of Tokyo, Tokyo 113-0033, Japan}
}
\date{}

\maketitle
\thispagestyle{empty}

\vspace{0cm}

\begin{abstract}
\normalsize
We study junctions consisting of confining strings in ${\cal N}=1$ supersymmetric
large $N$ gauge theories by means of the gauge/gravity correspondence.
We realize these junctions as D-brane configurations in infrared geometries of
the Klebanov-Strassler (KS) and the Maldacena-N\'u\~nez (MN) solutions.
After discussing kinematics associated with the balance of tensions,
we compute the energies of baryon vertices numerically.
In the KS background, baryon vertices give negative contributions to
the energies.
The results for the MN background strongly suggest that the energies
of baryon vertices exactly vanish,
as in the case of supersymmetric $(p,q)$-string junctions.
We find that brane configurations in the
MN background have a property similar to the holomorphy of
the M-theory realization of $(p,q)$-string junctions.
With the help of this property, we analytically prove the vanishing of
the energies of baryon vertices in the MN background.
\end{abstract}




\end{titlepage}
\section{Introduction}

The gauge/gravity correspondence\cite{maldacena,GKP,holography} relates
five-dimensional gravitational backgrounds
to four-dimensional field theories on the boundaries of the
five-dimensional spacetimes.
The extra dimension $r$ is related to the energy scales of the field theories
through red-shift (warp) factors depending on $r$.
The five-dimensional spacetimes are in general accompanied by compact
internal spaces.
The structures of these higher-dimensional spacetimes
reflect non-perturbative properties of their dual field theories.
Many gravity duals of various field theories
have been constructed in the context of string theory as near horizon geometries of branes
on which gauge theories are realized.

Various kinds of particles in boundary field theories are
identified with a number of objects in string theory.
These particles have been investigated with respect to the above-mentioned
duality.
Specifically, the spectra of glueballs\cite{thermal,grossooguri,glueball1,glueball2},
mesons\cite{KKW,mesonspec,sakaison,chiral,flavoring,excited,vectormeson,regge}
and (di-)baryons\cite{wittenbaryon,BfromSUGRA,imamura,callan,Bspec}
and interactions among them\cite{PS1,PS2,vectormeson}
have been studied with a variety of methods.
Bound states
of massive adjoint or bifundamental particles
are investigated in Refs. \citen{GZSS,ABC,KupSon,msspinning,BBCNZ,BCM,BCMZ}.
In Ref. \citen{penta}, the duality is used
to account for the extremely narrow decay width of
pentaquark baryons\cite{spring8,cern}.

\begin{figure}[htb]
\centerline{\includegraphics{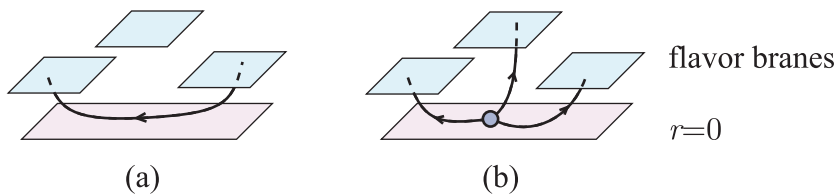}}
\caption{(a) A meson configuration and (b) a baryon configuration.}
\label{mesonbaryon.eps}
\end{figure}

Hadrons in $\SU(N)$ gauge theories are bound states of (anti-)quarks belonging to the (anti-)fundamental representation.
In order to introduce such quarks and antiquarks,
we need to place flavor D-branes in the dual gravity background\cite{adding}.
Quarks are identified with endpoints of open strings
on the flavor branes, and hadrons are constructed by
connecting them.
For example, a meson consisting of a quark and an antiquark
is realized as an open string
stretched between two flavor branes
on the gravity side,
as depicted in Fig. \ref{mesonbaryon.eps} (a).

Among the many states in a meson spectrum,
some low-lying states of pseudo-scalar and vector mesons are identified with
string modes representing fluctuations of the flavor branes and gauge fields on them.
The masses of such modes are examined in Refs. \citen{mesonspec,sakaison}, and the existence of localized modes with a mass gap is found in those works.

By contrast, Regge trajectories of higher-spin mesons can be treated
as semi-classical spinning strings in curved backgrounds\cite{mesonspec,regge}.
In general, it is difficult to determine the motion of strings completely,
and the Born-Oppenheimer approximation is often used to simplify such problems.
In this method, a quark-antiquark potential is
computed as the energy of a string with fixed endpoints\cite{reyyee,wl},
and then,
the motion of the endpoints (the quark and the antiquark) is
obtained by solving the resulting potential problem.
If the distance $L$ between a quark and an antiquark is large,
the $L$ dependence of the potential energy can in general be expanded as
\begin{equation}
E_{q\ol q}=TL+M_q+M_{\ol q}+{\cal O}(L^{-1}).
\label{qqbarexpansion}
\end{equation}
The coefficient $T$ of the first term here is the tension of the
confining string between the quark and the antiquark.
From the gravity point of view,
the first term is interpreted as the contribution
from the tension of a fundamental string at $r=0$,
because fundamental strings tend to approach $r=0$ as a result of
the gravitational force.
Thus, $T$ is obtained as the product of the redshift (warp) factor and the proper tension of fundamental strings.
In general, confining strings can be bound states of
elementary confining strings.
The tension of such strings depends non-linearly on the number of constituent elementary strings.
This property is reproduced in the gravity description\cite{tension} by taking account of
the effect of the transition of fundamental strings to D-branes expanded in internal spaces by Myers' effect\cite{myers}.
The second and third terms in (\ref{qqbarexpansion}),
which are independent of $L$, are interpreted as the self-energies of the quark and antiquark.
These can be evaluated by analyzing the catenary profile of a string near flavor D-branes.

A baryon is constructed by connecting flavor D-branes
with $N$ open fundamental strings of the same orientation.
In order to join such strings without violating the conservation law of the fundamental string charge,
we have to introduce a baryon vertex at the joint\cite{wittenbaryon,grossooguri}
(see Fig. \ref{mesonbaryon.eps} (b)).
This baryon vertex is a certain D-brane wrapped around a non-trivial cycle in the internal space.
In general, strings stretched between a baryon vertex and
a flavor D-brane could be bound,
and the number of confining strings meeting at a vertex can
be smaller than $N$.

Let us consider the potential among quarks in a baryon
in the spirit of the Born-Oppenheimer approximation.\cite{BfromSUGRA}
If the length of each confining string is large,
the energy of the baryon configuration is given by
\begin{equation}
E_{\rm baryon}=\sum_iT_iL_i+\sum_iM_i+E_{\rm vertex}+{\cal O}(L_i^{-1}),
\end{equation}
where $L_i$ and $T_i$ are the length and the tension of the $i$-th confining string,
and $M_i$ represents the self-energy of the quark at the end of the $i$-th confining string.
The third term, $E_{\rm vertex}$,
represents a new contribution which is absent in meson energies.
Naively, this contribution can be evaluated by multiplying the tension $T$
of the D-brane composing the baryon vertex,
the volume $V$ of the cycle around which the D-brane is wrapped, and the warp factor.
Although this does not give an exact result,
because of the deformation of the brane caused by the string tension\cite{imamura,callan},
we expect that a vertex gives some positive contribution of the order of $TV\times(\mbox{warp factor})$.
The purpose of this work is to compute this quantity using both
numerical and analytical methods.

This paper is organized as follows.
In the next section, we describe the context we consider
and clearly define the problem to be solved.
The junctions are realized as D3-brane configurations
embedded in a six-dimensional geometry.
In \S\ref{string.sec}, we briefly review the result of
Ref. \citen{tension} concerning the confining string tension
and study the kinematics of junctions associated with the tension balance.
In \S\ref{reduction.sec}, as preparation for the following computations,
we rewrite brane configurations
as two-dimensional surfaces in a certain four-dimensional target space
by assuming a certain symmetry.
The results of numerical computations are reported in \S\ref{numerical.sec},
and we present analytic results in \S\ref{analytic.sec}.
We conclude in \S\ref{conc.sec}.
In the appendix, we briefly explain how the electric fields on D3-branes are
treated in the numerical computations.
\section{D3-branes in the infrared geometry}\label{IR.sec}

We use two different classical solutions in type IIB supergravity
as background spacetimes dual to ${\cal N}=1$ supersymmetric confining gauge theories.
One is the Maldacena-N\'u\~nez (MN) solution\cite{MN1,MN2},
and the other is the Klebanov-Strassler (KS) solution\cite{kn,KS,lecture}.

The MN solution is the near horizon geometry
of $N$ coincident D5-branes wrapped around a two-cycle in a non-compact Calabi-Yau
manifold.
This is believed to be the gravity dual to the ${\cal N}=1$ $\SU(N)$ pure
Yang-Mills theory.
This theory has only two parameters, the size of
the gauge group, $N$, and the confinement scale, $\Lambda_{\rm QCD}$.

The KS solution can be interpreted as the near horizon geometry of
$N+k$ coincident D5-branes and $k$ anti-D5-branes wrapped around a two-cycle.
The corresponding boundary field theory is an ${\cal N}=1$
$\SU(N+k)\times\SU(k)$ gauge theory with bifundamental chiral multiplets.
The number $k$ depends on the energy scale.
As the energy decreases, $k$ also decreases due to cascade phenomenon\cite{KS}.
In the brane picture, this can be regarded as a kind of
pair annihilation of D5-branes and anti-D5-branes.
At very low energy, all the anti-D5-branes disappear with
the same number of D5-branes, leaving $N$ D5-branes.
Thus, the boundary theory of the KS solution
behaves like the ${\cal N}=1$ $\SU(N)$ supersymmetric Yang-Mills theory
in the low energy limit.
In addition to $N$ and $\Lambda_{\rm QCD}$, this theory includes the parameter $\ol g$, defined by $\ol g^{-2}=g_1^{-2}+g_2^{-2}$,
where $g_1$ and $g_2$ are the gauge coupling constants of $\SU(N+k)$ and
$\SU(k)$, respectively.
Although $g_1$ and $g_2$ run with the energy scale individually, $\ol g$ is
a renormalization invariant parameter.
This parameter is related to the string coupling constant $g_{\rm str}$,
which is constant in the KS solution, as
\begin{equation}
\ol g^2=4\pi g_{\rm str}.
\label{gbar}
\end{equation}

Corresponding to the similarity of the types of low-energy behavior of
the two gauge theories,
the two classical solutions have similar structure.
They are warped products of a five-dimensional manifold with coordinates $(x^\mu,r)$ and
an internal space with the topology ${\bf S}^2\times{\bf S}^3$.
${\bf S}^2$ shrinks to a point at $r=0$, while ${\bf S}^3$
possesses non-zero size everywhere.
The infrared (IR) dynamics of the field theories
are reflected by the structure near the centers of the classical solutions.
For both the MN and the KS solutions,
the metric
of the $r=0$ subspace relevant to the IR dynamics is given by
\begin{equation}
ds^2
=R^2(\eta_{\mu\nu}dy^\mu dy^\nu+d\Omega_3^2),
\quad(\mu,\nu=0,1,2,3)
\label{metric0}
\end{equation}
where $d\Omega_3^2$ is the metric of the unit $3$-sphere.
We refer to this as IR geometry.
The size $N$ of the low-energy gauge group $\SU(N)$
of the boundary gauge theories
is determined by the R-R $3$-form flux flowing through ${\bf S}^3$.
In other words, the integral of the R-R $3$-form field strength $G_3$
over the non-trivial $3$ cycle gives $N$:
\begin{equation}
N=\oint_{{\bf S}^3}G_3.
\label{G3}
\end{equation}

The boundary theories of the MN and KS solutions include
only adjoint and bifundamental fields.
In order to introduce fundamental quarks
into these theories,
we need flavor D-branes.
We can use D5\cite{WangHu,flavoring},
D7\cite{adding,sakaison}, or D9-branes\cite{WangHu}
wrapped around appropriate cycles as flavor branes.
Hadrons are constructed by connecting these branes with fundamental strings
joined by baryon vertices.
We assume that the lengths of the strings are large and focus on the vicinity
of baryon vertices, which are located at the bottom ($r=0$)
of the classical solutions, due to the gravitational force.
We do not discuss the effect of the endpoints of strings on the flavor branes,
and the arguments given in this paper are independent of the choice of the
flavor branes.

In the cases of the MN and KS backgrounds,
baryon vertices are D3-branes wrapped around
${\bf S}^3$\cite{GK,kn,KT}, and
fundamental strings in the IR geometry
are expanded to D3-brane tubes with section ${\bf S}^2$\cite{tension},
due to the existence of the $G_3$ flux
through Myers' effect\cite{myers}.
Therefore,
the junctions of the confining strings
are dual to the D3-branes, with the electric flux on them embedded
in the IR geometry.

The magnetic flux on D3-branes carries the charge of D-strings, which
in the case of the KS solution have recently been identified with axionic strings\cite{axionic,axionic2}.
We do not consider them in this paper.

The action of a D3-brane in
the IR geometry described by the metric (\ref{metric0}) and the R-R flux (\ref{G3}) is
\begin{equation}
S=-T_{\rm D3}\int d^4\sigma\sqrt{-\det
\left(
g_{ab}
+\frac{2\pi}{T_{\rm str}}F_{ab}\right)}
+2\pi\int F_2\wedge C_2,\quad(a,b=0,1,2,3)
\label{action0}
\end{equation}
where $T_{\rm str}=1/(2\pi\alpha')$ and $T_{\rm D3}=1/((2\pi)^3\alpha'^2g_{\rm str})$
are the tensions of the fundamental strings and D3-branes, respectively.
Also, $C_2$ is the R-R two-form potential and $G_3=dC_2$.
When a confining string consisting of $n$ elementary strings is
realized as a D3-brane configuration
with the topology ${\bf R}^2\times{\bf S}^2$,
the integer $n$
is defined as the fundamental string charge carried by the
D3-brane.
The fundamental string current is derived from the action (\ref{action0})
by differentiating it with respect to the NS-NS two-form potential $B_2$.
(In the derivation of the current, $F_2$ should be regarded as the $B$-gauge invariant
field strength $dA_1+B_2$.
Once we have obtained the current, we set $B_2=0$.)
$n$ is obtained by integrating the current as
\begin{equation}
n=-\oint_X dS_aD^a+\oint_X C_2
 =-\oint_X dS_aD^a+\int_Y G_3,\quad
 (a=1,2,3)
\label{nD}
\end{equation}
where $X$ is a non-trivial $2$-cycle in the D3-brane worldvolume,
and $Y$ is a $3$-disk with boundary $\partial Y=X$.
The flux density $D^a$ in (\ref{nD}) is defined by
\begin{equation}
D^a=\frac{1}{2\pi}\frac{\delta S_{\rm BI}}{\delta F_{0a}}.
\quad(a=1,2,3)
\label{Ddef}
\end{equation}
Note that we use $S_{\rm BI}$, the Born-Infeld part of the action (\ref{action0}), to define $D^a$.
The flux density defined in this way is 
invariant under gauge transformations of the R-R $2$-form potential $C_2$.
Due to the ambiguity in the choice of $Y$ in ${\bf S}^3$,
$n$ is defined only mod $N$.
$n$ is a conserved charge and does not change under continuous deformations
of the brane.

We define a ``flux angle'' $\theta^{\rm f}$ by
\begin{equation}
\theta^{\rm f}=\pi\frac{n}{N},
\label{fluxangle}
\end{equation}
for later use.
We fix the mod-$\pi$ ambiguity of $\theta^{\rm f}$ by stipulating that $0<\theta^{\rm f}<\pi$.
Because the sum of charges $n$ of the strings meeting at a baryon vertex must be a multiple of $N$,
the sum of the corresponding flux angles is a multiple of $\pi$.
If there are three branches,
the sum can be $\pi$ or $2\pi$.
These two are essentially the same, due to the charge conjugation
$\theta^{\rm f}\rightarrow\pi-\theta^{\rm f}$.
If there are more than three branches
there are more than two genuinely different cases,
as we see below explicitly for four-string junctions.

To clarify the dependence of the action (\ref{action0}) on the parameters $R$ and $N$,
we factor these parameters out of the induced metric and
the R-R $3$-form flux,
writing them as
\begin{equation}
g_{ab}=R^2\wt g_{ab},\quad
G_3=\frac{N}{2\pi^2}\omega_3,
\end{equation}
where $\omega_3$ is the volume form of ${\bf S}^3$ normalized so that
its integral over ${\bf S}^3$ gives $\oint_{{\bf S}^3}\omega_3=2\pi^2$,
the volume of the unit $3$-sphere.
Next, we introduce a rescaled
field strength $\wt F_2$ and the corresponding potential
$\wt A_1$ as
\begin{equation}
F_2=\frac{R^2T_{\rm str}}{2\pi}\wt F_2,\quad
A_1=\frac{R^2T_{\rm str}}{2\pi}\wt A_1.
\end{equation}
The D3-brane action rewritten
in terms of these rescaled fields is
\begin{equation}
\wt S=-\int d^4\sigma\sqrt{-\det
\left(
\wt g_{ab}
+\wt F_{ab}\right)}
+\rho\int\wt A_1\wedge\omega_3,\quad
\rho=\frac{NT_{\rm str}}{2\pi^2 R^2T_{\rm D3}}.
\label{renS}
\end{equation}
Here, we have changed the normalization of the action
from $S$ to $\wt S$, which are related by $S=T_{\rm D3}R^4\wt S$.
This does not affect the classical equations of motion in which we are interested.
The dimensionless quantity $\rho$ is the unique parameter of this rescaled action.
It is related to the parameter $b$ used in Ref. \citen{tension}
by $\rho=2/b$.
Also note the relation
\begin{equation}
\frac{2}{\rho}=b=\frac{R^2}{R_{\rm D5}^2},
\label{rb}
\end{equation}
where $R_{\rm D5}=(N\alpha'g_{\rm str})^{1/2}$ is the radius of ${\bf S}^3$ near the horizon of
$N$ coincident flat D5-branes in the flat Minkowski background.
Because the MN solution can
be regarded as the near horizon geometry
of D5-branes wrapped around ${\bf S}^2$, the radius $R$ is identical to
$R_{\rm D5}$, and $b=1$.
However, for the KS solution, $b$
is modified due to the existence of anti D5-branes,
and its numerical value is $b=0.932660368\cdots$\cite{KS,tension}.
From the action (\ref{renS}), it is seen that we can regard the dimensionless
parameter $\rho$ as a charge density
coupled to the gauge field $\wt A_1$.

In this paper, we only consider static configurations, and
we always omit the time dimension in the following.
By the term ``worldvolume'' we refer to a time slice
of the entire worldvolume.

Instead of $D^a$ defined by (\ref{Ddef}),
we use the rescaled flux density and its Hodge dual defined by
\begin{equation}
\wt D^a=\frac{\delta\wt S}{\delta\wt F_{0a}}=\frac{2\pi^2\rho}{N}D^a,\quad
\wt D_2=\frac{1}{2}\sqrt{\det\wt g_{ab}}\epsilon_{abc}D^a d\sigma^b\wedge d\sigma^c.\quad
(a,b,c=1,2,3)
\end{equation}
The relation (\ref{nD}) rewritten in terms of the flux angle and the rescaled flux density becomes
\begin{equation}
2\pi\rho\theta^{\rm f}=-\oint_X\wt D_2+\rho\int_Y \omega_3.
\label{invflux}
\end{equation}
This is the integral form of the Gauss's Law constraint,
\begin{equation}
d\wt D_2=\rho\omega_3|_{\rm brane},
\label{gauss}
\end{equation}
where $\omega_3|_{\rm brane}$ is the pull-back of $\omega_3$ to the D3-brane worldvolume.

The energy of a D3-brane is given by $E=T_{\rm D3}R^3\wt E$ where
$\wt E$ is the rescaled energy
\begin{equation}
\wt E=\int d^3\sigma\sqrt{\det\wt g_{ab}}\sqrt{1+\wt g_{ab}\wt D^a\wt D^b}.\quad
(a,b=1,2,3)
\label{energy}
\end{equation}
The shape of a D3-brane and the electric flux density on it should be determined so that energy is minimized.

In the following sections, we use only rescaled
quantities, such as $\wt E$, $\wt D^a$ and $\wt g_{ab}$.
In the rest of this section, we summarize the relations between these dimensionless
quantities and dimensionful quantities in boundary field theories.
From the definition of the rescaled quantities $\wt S$ and $\wt g_{ab}$,
we can determine the relations between the rescaled energies and tensions
and original ones.
To determine the corresponding quantities in boundary field theories,
we should also take account of the warp factor $W$, which determines
the ratio of the energies in the IR geometry and those in the boundary field theories.
Doing this, we obtain
\begin{eqnarray}
E_{\rm vertex}^{(\rm FT)}&=&WT_{D3}R^3\wt E=\left(\frac{W}{\alpha'^{1/2}}\right)\frac{bN^{3/2}g_{\rm str}^{1/2}}{(2\pi)^3}\wt E_{\rm vertex},\label{EFT}\\
T^{(\rm FT)}
&=&W^2T_{D3}R^2\wt T=\left(\frac{W}{\alpha'^{1/2}}\right)^2\frac{bN}{(2\pi)^3}\wt T,
\end{eqnarray}
where $\wt E_{\rm vertex}$ and $\wt T$ are the energy of a baryon vertex and a confining string tension,
given in the following sections, and $E^{(\rm FT)}_{\rm vertex}$ and $T^{(\rm FT)}$ are
the corresponding quantities in the boundary field theories.

The confinement scale $\Lambda_{\rm QCD}$ in the boundary gauge theories can be defined
by $T^{(\rm FT)}\sim\Lambda_{\rm QCD}^2$.
Thus, up to a numerical factor, we have
\begin{equation}
\Lambda_{\rm QCD}\sim\frac{W}{\alpha'^{1/2}}.
\end{equation}
(Here we have used the tension of an elementary confining string, $\wt T\sim N^{-1}$.)
Then, the energy of a baryon vertex can be rewritten as
\begin{equation}
E_{\rm vertex}^{(\rm FT)}\sim\Lambda_{\rm QCD}N^{3/2}g_{\rm str}^{1/2}\wt E_{\rm vertex}.
\label{EFT2}
\end{equation}
This depends on the string coupling constant, $g_{\rm str}$.
In the KS case, $g_{\rm str}$ is related to the parameter $\ol g$ by (\ref{gbar}),
and we obtain the following relation, which includes only field theory variables:
\begin{equation}
E^{(\rm FT)}_{\rm vertex}\sim\Lambda_{\rm QCD}N^{3/2}\ol g\wt E_{\rm vertex}.
\label{EFT3}
\end{equation}
By contrast,, there is no parameter corresponding to $g_{\rm str}$ in the MN case.
We return to this problem in \S\ref{analytic.sec}.

In the following sections, we use only dimensionless quantities,
and we omit the tildes on rescaled variables for simplicity.
\section{Confining strings and their junctions}\label{string.sec}
In this section we first briefly review how the tensions of confining strings are computed
as the energies of D3-branes, following Ref. \citen{tension}.
Next, we study the kinematics of junctions by considering the balance of tensions on vertices.

Let $(x,y,z)$ be the three spatial coordinates of the boundary.
Combining these with the coordinates of the internal space ${\bf S}^3$,
we have the set of coordinates
$(x,y,z,\theta,\phi,\psi)$ with the following metric:
\begin{equation}
ds^2=dx^2+dy^2+dz^2+d\theta^2+\sin^2\theta(d\phi^2+\sin^2\phi d\psi^2).
\label{xytpp}
\end{equation}
The ranges of the angular coordinates are
\begin{equation}
0\leq\theta\leq\pi,\quad
0\leq\phi\leq\pi,\quad
0\leq\psi<2\pi.
\end{equation}
With this parameterization,
the volume form of ${\bf S}^3$ is given by
\begin{equation}
\omega_3=\sin^2\theta\sin\phi d\theta\wedge d\phi\wedge d\psi.
\end{equation}

The dual configuration of an infinitely long confining string
is a D3-brane with the worldvolume ${\bf S}^2\times{\bf R}$.
We begin with the ansatz
\begin{equation}
x=\sigma^1,\quad
y=0,\quad
z=0,\quad
\theta=\theta^{\rm r},\quad
\phi=\sigma^2,\quad
\psi=\sigma^3.
\end{equation}
The parameter $\theta^{\rm r}$ is a constant representing the angular radius
of ${\bf S}^2\subset{\bf S}^3$.
By rotational symmetry, we
can easily determine the flux density $D^a$ as a function of $\theta^{\rm f}$ and $\theta^{\rm r}$
from (\ref{invflux}).
Its only non-vanishing component is $D^x$, and it is given explicitly by
\begin{equation}
D^x=\frac{-1}{b\sin^2\theta^{\rm r}}[\theta^{\rm f}- (\theta^{\rm r}-\sin\theta^{\rm r}\cos\theta^{\rm r})].
\end{equation}
Substituting this into (\ref{energy}), we obtain
the tension of a confining string.
\begin{equation}
T=\frac{dE}{dx}=4\pi\sqrt{\sin^4\theta^{\rm r}+\frac{1}{b^2}(\theta^{\rm f}-\theta^{\rm r}+\sin\theta^{\rm r}\cos\theta^{\rm r})^2}.
\label{eoverl}
\end{equation}
We should determine the angle $\theta^{\rm r}$ as the point of minimum tension.
The condition $dT/d\theta^{\rm r}=0$ gives the relation between
$\theta^{\rm f}$ and  $\theta^{\rm r}$,
\begin{equation}
\theta^{\rm r}-(1-b^2)\sin\theta^{\rm r}\cos\theta^{\rm r}=\theta^{\rm f}.
\label{mincond}
\end{equation}
If $0\leq b^2\leq 1$, this relation defines a one to one mapping between
$0\leq\theta^{\rm f}\leq\pi$ and $0\leq\theta^{\rm r}\leq\pi$.
The minimum value of the tension is
\begin{equation}
T=4\pi\sin\theta^{\rm r}\sqrt{\sin^2\theta^{\rm r}+b^2\cos^2\theta^{\rm r}}.
\label{tensionformula}
\end{equation}

For the MN solution,
(\ref{mincond}) can be solved immediately, and
the tension (\ref{tensionformula}) reduces to the simple form
\begin{equation}
\theta^{\rm f}=\theta^{\rm r},\quad
T=4\pi\sin\theta^{\rm f},\quad
\mbox{for $b=1$}.
\label{ttheta}
\end{equation}
Contrastingly, the angle $\theta^{\rm r}$ and the tension for the KS solution can
only be obtained numerically.
The tensions $T$ and angles $\theta^{\rm r}$ for several values of $\theta^{\rm f}$ are listed in Table \ref{tensions.tbl}.
These values are used below to compute the energies of the baryon vertices.
\begin{table}[htb]
\caption{The angle $\theta^{\rm r}$ and the tension $T$ for several values of $\theta^{\rm f}$}
\label{tensions.tbl}
\begin{center}
\begin{tabular}{r|rr|rr}
\hline
\hline
&
\multicolumn{2}{c|}{MN ($b=1$)}&
\multicolumn{2}{c}{KS ($b=0.9326604$)} \\
\multicolumn{1}{c|}{$\theta^{\rm f}$} &
\multicolumn{1}{c}{$\theta^{\rm r}$} &
\multicolumn{1}{c|}{$T$} &
\multicolumn{1}{c}{$\theta^{\rm r}$} &
\multicolumn{1}{c}{$T$} \\
\hline
 $\pi/12=0.261799$ & $0.261799$ &  $3.252416$ & $0.298366$ &  $3.467446$ \\
$2\pi/12=0.523599$ & $0.523599$ &  $6.283185$ & $0.583434$ &  $6.601513$ \\
$3\pi/12=0.785398$ & $0.785398$ &  $8.885766$ & $0.849929$ &  $9.168758$ \\
$4\pi/12=1.047198$ & $1.047198$ & $10.882796$ & $1.099822$ & $11.047175$ \\
$5\pi/12=1.308997$ & $1.308997$ & $12.138181$ & $1.338189$ & $12.185588$ \\
$6\pi/12=1.570796$ & $1.570796$ & $12.566371$ & $1.570796$ & $12.566371$ \\
\hline
\end{tabular}
\end{center}
\end{table}

In the cases of both the MN and KS backgrounds, the tension
depends non-linearly on $\theta^{\rm f}$.
This non-linear dependence implies the formation of
truly bound states and the absence of supersymmetry.
Therefore we cannot apply the method employing Killing spinors,
which is useful to determine the baryon configuration in ${\cal N}=4$ Yang-Mills
theory\cite{imamura}, to configurations investigated in this paper.

Using the tension formula obtained above,
we now consider the kinematics of three- and four-string junctions.
We first study three-string junctions.
The angles at which the strings meet are fixed by the requirement that
of the tensions balance at the vertex.
\begin{figure}[htb]
\centerline{\includegraphics{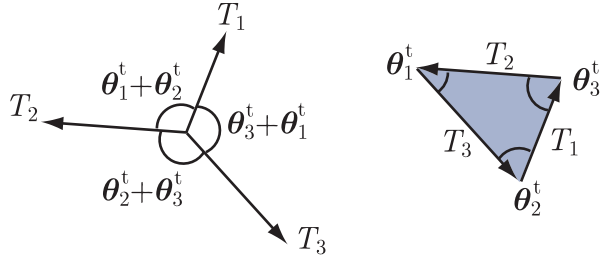}}
\caption{The balance of three tensions.}
\label{balance.eps}
\end{figure}
If we define the angle $\theta^{\rm t}_i$ as the angle opposite
the force vector of the $i$-th string in the triangle consisting of
three force vectors,
the angle made by the $i$-th and $j$-th strings is $\theta_i^{\rm t}+\theta_j^{\rm t}$
(see Fig. \ref{balance.eps}).
If the tensions of the three strings are $T_i$, $i=1,2,3$,
the angles are uniquely determined by
\begin{equation}
\theta_1^{\rm t}=\cos^{-1}\frac{T_2^2+T_3^2-T_1^2}{2T_2T_3},\quad
0<\theta_1^{\rm t}<\pi,
\end{equation}
and similar equations for $\theta_2^{\rm t}$ and $\theta_3^{\rm t}$.
These are equivalent to the following:
\begin{equation}
T_1:T_2:T_3=\sin\theta_1^{\rm t}:\sin\theta_2^{\rm t}:\sin\theta_3^{\rm t},\quad
\theta^{\rm t}_1+\theta^{\rm t}_2+\theta^{\rm t}_3=\pi.
\label{t1t2t3}
\end{equation}
Unlike the relation between $\theta_i^{\rm f}$ and $\theta_i^{\rm r}$,
$\theta_i^{\rm t}$ is not a function of only the single angle $\theta_i^{\rm f}$ with the same index $i$.
Rather, it is a function of the set of angles $\{\theta_1^{\rm f},\theta_2^{\rm f},\theta_3^{\rm f}\}$.
Because the sum of the three flux angles is $\pi$ or $2\pi$,
only two of them are independent.
We represent $\theta_1^{\rm t}$ as a function of $\theta_2^{\rm f}$ and $\theta_3^{\rm f}$
and denote it by $\theta_1^{\rm t}=\tau(\theta_2^{\rm f},\theta_3^{\rm f})$.
The other two angles are similarly given by
$\theta_2^{\rm t}=\tau(\theta_3^{\rm f},\theta_1^{\rm f})$
and $\theta_3^{\rm t}=\tau(\theta_1^{\rm f},\theta_2^{\rm f})$,
with the same function $\tau$.

For the MN case, (\ref{t1t2t3}) reduces to the simple relation $\sin\theta_i^{\rm t}=\sin\theta_i^{\rm f}$.
Even in this case, we cannot determine
$\theta_i^{\rm t}$ from only the single angle $\theta_i^{\rm f}$,
because there are two solutions, $\theta_i^{\rm t}=\theta_i^{\rm f}$
and $\theta_i^{\rm t}=\pi-\theta_i^{\rm f}$.
These two solutions correspond to the two possible values $\pi$ and $2\pi$ of
the total flux angle, $\theta_{\rm tot}^{\rm f}=\sum_{i=1}^3\theta_i^{\rm f}$, respectively.
These two cases can be expressed together by the single equation
\begin{equation}
\theta_3^{\rm t}=\tau(\theta_1^{\rm f},\theta_2^{\rm f})=|\pi-\theta_1^{\rm f}-\theta_2^{\rm f}|.
\end{equation}

\begin{figure}[htb]
\centerline{\includegraphics{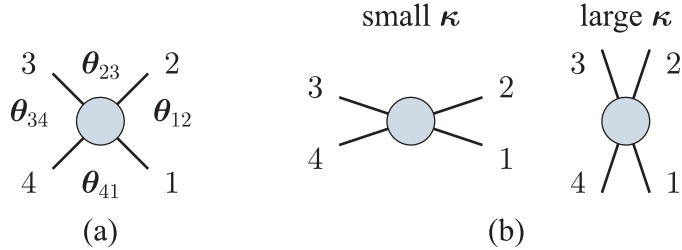}}
\caption{A junction with four external strings.}
\label{fourjunc.eps}
\end{figure}

We now proceed to consider four-string junctions.
We assume that the four external strings are on a plane.
We label the external strings in
counterclockwise order by $1$, $2$, $3$ and $4$,
and denote the angle made by strings $i$ and $i+1$
by $\theta_{i,i+1}$.
[see Fig. \ref{fourjunc.eps} (a)]
(The labels of the strings are defined mod 4.)
The sum of these four angles is $2\pi$, and only three of them are independent.
If the flux angles $\{\theta_1^{\rm f},\theta_2^{\rm f},\theta_3^{\rm f},\theta_4^{\rm f}\}$
of the four branches
are given, the tensions are determined by the formula above,
and the balance of these four tensions imposes two conditions on the angles $\theta_{i,i+1}$.
These conditions leave one degree of freedom unfixed.
We define a parameter $\kappa$ parameterizing this unfixed degree of freedom.
One useful choice is
\begin{equation}
\kappa=\theta_{12}+\theta_{34}=2\pi-(\theta_{23}+\theta_{41}).
\label{kappadef}
\end{equation}
If this parameter changes, the junction deforms as in Fig. \ref{fourjunc.eps}(b).

\begin{figure}[htb]
\centerline{\includegraphics{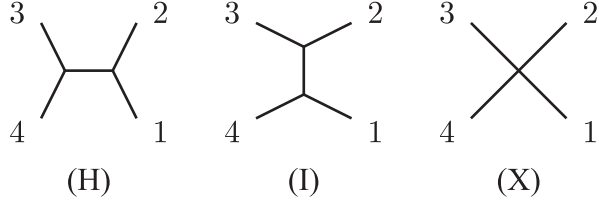}}
\caption{The three possible topologies of four-string junctions.
Two of them include two three-string vertices and the other has
one four-string vertex.}
\label{topology.eps}
\end{figure}

There are three possible topologies of four-string junctions, as shown in
Fig. \ref{topology.eps}.
For junctions with the topologies (H) and (I), which include two three-string vertices,
the angles $\theta_{i,i+1}$ are uniquely determined by
the flux angles, and the corresponding $\kappa$-parameters are given by
\begin{equation}
\kappa_H=2\pi-\tau(\theta_1^{\rm f},\theta_2^{\rm f})-\tau(\theta_3^{\rm f},\theta_4^{\rm f}),\quad
\kappa_I=\tau(\theta_2^{\rm f},\theta_3^{\rm f})+\tau(\theta_4^{\rm f},\theta_1^{\rm f}).
\label{thetaiip1}
\end{equation}

Let us suppose that there is a junction with the topology (H).
This can be in equilibrium if $\kappa=\kappa_H$.
If we try to change the parameter $\kappa$ by changing the directions of the strings,
the two vertices move in such a way that the movement compensates for the variation of $\kappa$.
If we attempt to decrease $\kappa$, two vertices move away from each other,
with the result that $\kappa$ remains unchanged.
If we move the strings in the opposite direction,
the two vertices approach each other.
This compensating mechanism consisting of the movement of vertices is effective until the two vertices coincide.
If we continue to move the strings after the vertices come to coincide,
the topology of the junction tends to change to (I) via (X).
Until the instant at which the two vertices first coincide,
we have $\kappa=\kappa_H$, and
the behavior of the junction after this instant depends on the relation between
$\kappa_H$ and $\kappa_I$.

If $\kappa_H>\kappa_I$, just as the two vertices of (H) meet,
the topology of the junction changes to (I)
and becomes unstable,
and then the two vertices separate rapidly, eventually realizing
the equilibrium condition, $\kappa=\kappa_I$.
In this case, the junctions of the topology (X) are unstable.

If $\kappa_H<\kappa_I$, even after the two vertices coincide,
the topology does not change to (I), and (X) is stable until
$\kappa$ reaches $\kappa_I$.
When $\kappa$ reaches $\kappa_I$, the topology of the junction finally changes to (I).

When $\kappa_H=\kappa_I$, there is a unique value of $\kappa$ that gives stable configurations,
and the three topologies can change to each other marginally without changing $\kappa$.

To express these types of behavior of the junctions, it is convenient to define
the parameter $\lambda\equiv\kappa_H-\kappa_I$ representing
the ``repulsive force'' between two vertices.
If $\lambda$ is positive, two three-string vertices repel each other,
and they cannot merge to form a stable four-string vertex.
If $\lambda$ is negative, two vertices attract each other,
and they are bound
into a four-string vertex, provided that $\kappa_H<\kappa<\kappa_I$.

Which cases are realized for junctions
in the MN and KS backgrounds?
From (\ref{kappadef}) and (\ref{thetaiip1}),
$\lambda=\kappa_H-\kappa_I$ is given by
\begin{equation}
\lambda
=2\pi
-\tau(\theta_1^{\rm f},\theta_2^{\rm f})
-\tau(\theta_2^{\rm f},\theta_3^{\rm f})
-\tau(\theta_3^{\rm f},\theta_4^{\rm f})
-\tau(\theta_4^{\rm f},\theta_1^{\rm f}).
\end{equation}

For the MN solution, the explicit form of $\lambda$ is
\begin{equation}
\lambda
=2\pi
-|\pi-\theta_1^{\rm f}-\theta_2^{\rm f}|
-|\pi-\theta_2^{\rm f}-\theta_3^{\rm f}|
-|\pi-\theta_3^{\rm f}-\theta_4^{\rm f}|
-|\pi-\theta_4^{\rm f}-\theta_1^{\rm f}|.
\end{equation}
We can easily show that $\lambda=0$ when $\sum_{i=1}^4\theta_i^{\rm f}=\pi$ or $\sum_{i=1}^4\theta_i^{\rm f}=3\pi$
and that $\lambda>0$ when $\sum_{i=1}^4\theta_i^{\rm f}=2\pi$.

For the KS solution, although it is difficult to determine the signature of $\lambda$ analytically,
we can numerically check that it is always positive.
This implies that planar four-string junctions are always unstable in the KS background.

\section{Reduction to lower dimensions}\label{reduction.sec}
In the last section, we saw that
junctions are described as three-dimensional surfaces in the six-dimensional space spanned by the
coordinates $(x,y,z,\theta,\phi,\psi)$.
We only consider planar junctions and set $z=0$.
If we ignore the extension along the $(x,y)$-plane,
the worldvolume of a D3-brane representing
a $k$-string junction is ${\bf S}^3$ with $k$ punctures.
Each puncture is topologically a three-dimensional disc
and corresponds to each branch string.
Let us assume that these punctures are located on a large circle
of ${\bf S}^3$.
In this case, the configuration possesses $\U(1)$ symmetry,
as determined by the symmetry group of this circle.
It is convenient to choose ${\bf S}^1$ given by $g_{\psi\psi}=0$ as the
fixed large circle so that
the $\U(1)$ action constitutes a constant shift of the coordinate $\psi$.
General brane configurations with this $\U(1)$ symmetry are
\begin{equation}
x=x(\sigma^1,\sigma^2),\quad
y=y(\sigma^1,\sigma^2),\quad
\theta=\theta(\sigma^1,\sigma^2),\quad
\phi=\phi(\sigma^1,\sigma^2),\quad
\psi=\sigma^3.
\end{equation}
Thus, we can represent configurations as two-dimensional surfaces
in the four-dimensional space with the coordinates $(x,y,\theta,\phi)$.

\begin{figure}[htb]
\centerline{\includegraphics{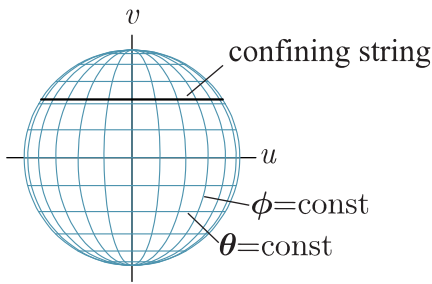}}
\caption{The projection of the hemisphere parameterized by $(\theta,\phi)$
onto the unit disc in
the $(u,v)$-plane.
The cross-section of a confining string solution is represented as a chord
of this unit circle in the $(u,v)$-plane.}
\label{proj.eps}
\end{figure}

Instead of the angular coordinates $(\theta,\phi)$,
it is convenient to use $(u,v,w)$ satisfying
constraints $u^2+v^2+w^2=1$ and $w\geq0$.
The relations among these coordinates are
\begin{equation}
u=\sin\theta\cos\phi,\quad
v=\cos\theta,\quad
w=\sin\theta\sin\phi
\end{equation}
(see Fig. \ref{proj.eps}).
If we use $(x,y,u,v)$ as a set of independent coordinates,
the target space is ${\bf R}^2\times{\bf D}^2$, where ${\bf R}^2$ is
spanned by $x$ and $y$, and ${\bf D}^2$ is the unit disc satisfying
$u^2+v^2\leq1$ in the $(u,v)$-plane.
The variable $w=\sqrt{1-u^2-v^2}$ is treated as a function of
$u$ and $v$.
Brane configurations are two-dimensional surfaces
embedded in this four-dimensional target space.
For example, a confining string solution given in \S\ref{string.sec}
is represented as a band in the four-dimensional space which is a direct product of a chord $v=\cos\theta^{\rm r}$ on the unit disc
and an infinitely long line in the $(x,y)$-plane.

We should redefine the electric flux density as a field on the two-dimensional surface
by integrating the flux density $D_2$ over $\psi$.
We introduce a one-form $D_1'$ and
its dual vector $D'^a$ as
\begin{equation}
D_1'=\int_\psi D_2,\quad
D_1'=\sqrt{\det g_{ab}}\epsilon_{ab}D'^ad\sigma^b.
\quad(a,b=1,2)
\end{equation}
The Gauss's Law constraint satisfied by $D_1'$ is
obtained by integrating the constraint (\ref{gauss})
given by
\begin{equation}
dD'_1=2\pi\rho du\wedge dv.
\label{d1gauss}
\end{equation}
The corresponding integral form is
\begin{equation}
2\pi\rho\theta^{\rm f}=-\oint_C D'_1+2\pi\rho\int_A du\wedge dv,
\label{invflux2}
\end{equation}
where $C$ and $A$
correspond to $X$ and $Y$ in (\ref{invflux}), respectively.
$C$ is a curve on the surface connecting two boundaries.
$A$ is a two-dimensional surface whose boundary consists of
$C$ and a curve in the boundary $u^2+v^2=1$ of the four-dimensional target space.
To obtain the right-hand sides of Eqs.
(\ref{d1gauss}) and (\ref{invflux2}), we have used the relation
\begin{equation}
\int_\psi\omega_3=2\pi\sin^2\theta\sin\phi d\theta\wedge d\phi=2\pi du\wedge dv.
\label{rhoprime}
\end{equation}
The energy (\ref{energy}) integrated over $\psi$ is
\begin{equation}
E=\int d^2\sigma\sqrt{\det g_{ab}}\sqrt{(2\pi w)^2+ g_{ab}D'^aD'^b}.
\quad (a,b=1,2)
\label{2dime}
\end{equation}

Now, the problem that we wish to solve is
to find a two-dimensional surface in the four-dimensional space $(x,y,u,v)$
and a flux density $D_1'$ on it
which minimize the energy
(\ref{2dime}) under the constraint (\ref{invflux2})
imposed on each external confining string.

\begin{figure}[htb]
\centerline{\includegraphics{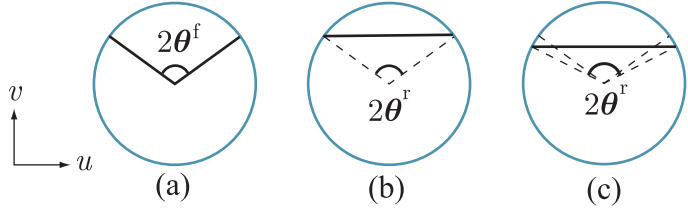}}
\caption{The projections of brane configurations in the $(u,v)$-plane.
(a) The wedge configuration used as an initial configuration.
(b), (c) The confining string solutions in the MN and KS backgrounds, respectively.}
\label{wmnks.eps}
\end{figure}

When we perform a numerical computation to determine
a brane configuration, we start from an initial configuration
and look for a configuration with minimum energy by varying the
shape of the surface and the flux on it.
We can choose any configuration as the initial configuration, as long as it satisfies
appropriate boundary conditions and (\ref{invflux2}),
and as long as it is in the same topological class as
the final configuration we wish to obtain.
One natural choice for a confining string is a D3-brane with a shrunk worldvolume, for which the second term of (\ref{invflux}) vanishes.
This represents a brane before being expanded by Myers' effect.
This choice, however, is very singular and not appropriate for the numerical computation.
Therefore, we instead adopt a worldvolume on which the electric flux density vanishes everywhere on the
surface.
In this case, the constraint (\ref{invflux2}) requires that the area in the $(u,v)$-plane
enclosed by $C$ and the unit circle $u^2+v^2=1$ be the same as the flux angle $\theta^{\rm f}$.
The most convenient one satisfying this condition
is the ``wedge configuration''
defined as a direct product of a line in the $(x,y)$-plane
and a wedge in the $(u,v)$-plane, as shown in Fig. \ref{wmnks.eps} (a).

Starting from a wedge configuration with a central angle $2\theta^{\rm f}$,
we can numerically reproduce the corresponding confining string solution
represented as a chord in the $(u,v)$-plane.
Its length is determined by $\theta^{\rm r}$ defined in \S\ref{string.sec}.
For the MN solution,
the two angles $\theta^{\rm f}$ and $\theta^{\rm r}$ are equal.
This implies that the distance between the two endpoints
of the wedge on the unit circle is the same as the length of
the chord
[see Fig. \ref{wmnks.eps} (b)].
However,
this distance becomes slightly greater for the KS solution.
[see Fig. \ref{wmnks.eps} (c)].

\begin{figure}[htb]
\centerline{\includegraphics{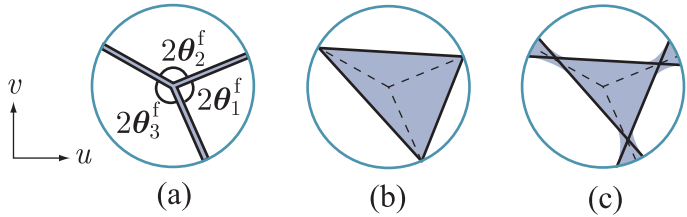}}
\caption{(a) An initial configuration for a three string junction
is composed of three wedge configurations.
After minimizing the energy by varying the configuration,
we obtain the junction solutions for (b) the MN background
and (c) the KS background, respectively.}
\label{baryonuv.eps}
\end{figure}

Wedge configurations are also suitable to construct initial configurations for junctions,
because they can be pasted just like
interaction vertices in Witten's open string field theory\cite{cubic},
as shown in
Fig. \ref{baryonuv.eps} (a).
By varying this initial configuration and minimizing the energy,
we obtain a junction configuration.

In the case of the MN background,
the distance between the endpoints of
each curve representing each branch string
does not change, and
the junction solution is expected to be a triangle inscribed in the unit circle
in the $(u,v)$-plane
[see Fig. \ref{baryonuv.eps} (b)].
Indeed, the coincidence of the endpoints of the three chords is guaranteed by
the relations (\ref{modCR}),
given below.
It is important that this triangle is similar to the tension triangle in Fig. \ref{balance.eps}.

For the KS solution, the chords representing asymptotic string solutions are too long
to make an inscribed triangle
[see Fig. \ref{baryonuv.eps} (c)].

\section{Numerical analysis}\label{numerical.sec}
In this section, we report our results of our numerical study of brane configurations.
We start by reconstructing the confining string solutions
to assess the accuracy of the results of the numerical method
by comparing them to the analytic results given in \S\ref{string.sec}.

\begin{figure}[htb]
\centerline{\includegraphics[width=.8\linewidth]{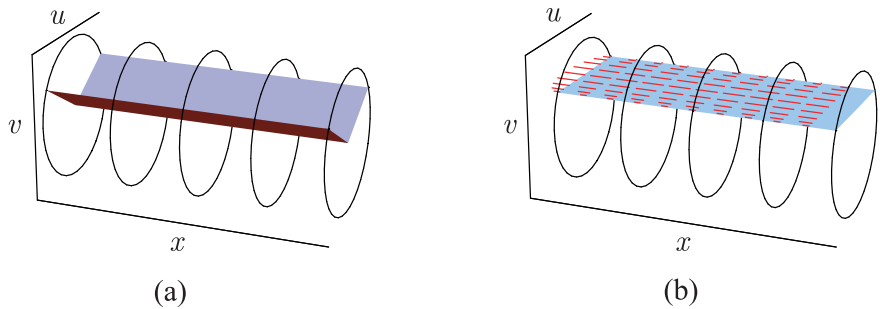}}
\caption{(a) An initial wedge configuration and (b) a numerically generated confining string solution.
The segments in the worldvolume in (b) represent the flux density.}
\label{wedge.eps}
\end{figure}

\begin{figure}[htb]
\centerline{\includegraphics{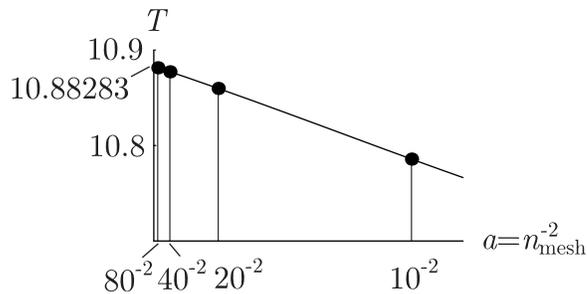}}
\caption{A fitting of numerically generated tensions for several values $n_{\rm mesh}$.}
\label{extra.eps}
\end{figure}

We first prepare a wedge configuration with a central angle $2\theta^{\rm f}$ and a length $L$ as an initial
configuration [see Fig. \ref{wedge.eps} (a)].
We compute the energy of the surface by means of the triangulation method.
The surface is divided into a mesh of $2n_{\rm mesh}^2$ small triangular regions.
The shape of the surface is represented by the positions of the sites, and
the flux density is treated as link variables.
We seek a configuration that minimizes the energy by varying these variables
while enforcing the Gauss's Law constraint.
The detailed algorithm is given in the appendix.
A result for ($\theta^{\rm f}=\pi/3$, $L=4$, $b=1$) is given in Table \ref{tension.tbl}.
\begin{table}[htb]
\caption{A numerical result for the tension of a confining string with $\theta^{\rm f}=\pi/3$
in the MN solution.
The value for $n_{\rm mesh}\rightarrow\infty$ is obtained by
extrapolating the data for $n_{\rm mesh}=10$, $20$, $40$ and $80$
with the fitting function $c_0+c_1n_{\rm mesh}^{-2}+c_2n_{\rm mesh}^{-4}+c_3n_{\rm mesh}^{-6}$.
}
\label{tension.tbl}
\begin{center}
\begin{tabular}{c|rrrrr|c}
\hline
\hline
$n_{\rm mesh}$   &
\multicolumn{1}{c}{$10$} &
\multicolumn{1}{c}{$20$} &
\multicolumn{1}{c}{$40$} &
\multicolumn{1}{c}{$80$} &
\multicolumn{1}{c|}{$\infty$} & exact \\
\hline
$T=E/L$ & $10.7858$ & $10.85978$ & $10.87722$ & $10.88144$ & $10.88283$ & $10.88280$ \\
error & $-0.0970$ & $-0.02302$ & $-0.00558$ & $-0.00136$ & $+0.00003$ & $-$ \\
\hline
\end{tabular}
\end{center}
\end{table}
By comparing this result to the corresponding tension in Table \ref{tensions.tbl},
we can estimate the
accuracy of the outputs of the numerical analysis.
We find a relative error of order $\sim10^{-4}$
for the finest meshes, with $n_{\rm mesh}\sim 80$.
Plotting the error as a function of
the typical size of triangles, $a=n_{\rm mesh}^{-2}$,
we find that the error is almost proportional to $a$
(see Fig. \ref{extra.eps}).
The value for $n_{\rm mesh}=\infty$ ($a=0$) given in Table \ref{tension.tbl}
is obtained by extrapolating the data for finite $n_{\rm mesh}$ with a polynomial of $a$.
This extrapolation gives a value very close to the
exact one, with a relative error $<10^{-5}$.
Because of this small error, we use the same extrapolation to obtain the energies
of the baryon vertices.

\begin{figure}[htb]
\centerline{\includegraphics[width=.4\linewidth]{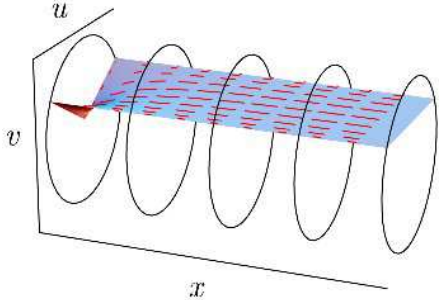}}
\caption{The brane configuration with a boundary on which the fixed boundary condition is imposed.}
\label{dump.eps}
\end{figure}

Let $E_{\rm junc}(L)$ be the energy of a three string junction with the branch length $L$.
Then, the contribution from a baryon vertex is defined by
\begin{equation}
E_{\rm vertex}=\lim_{L\rightarrow\infty}\Big(E_{\rm junc}(\theta_i^{\rm f},L)-L\sum_iT(\theta_i^f)\Big).
\label{ebdef}
\end{equation}
However,
we cannot in practice numerically compute $E_{\rm junc}$ in the limit $L\rightarrow\infty$.
Therefore, instead, we use a sufficiently long fixed length
beyond which the effect of the vertices becomes negligible.
In the following way,
we find evidence that $L=4$ is sufficiently large to
to guarantee a relative accuracy of $10^{-5}$.
First, we compute the energies of two strings stretched between $x=0$ and $x=L$ with
lengths $L=4$ and $L=8$.
The worldvolume of the short and the long strings are composed of $2n_{\rm mesh}^2$ and
$4n_{\rm mesh}^2$ triangles, respectively.
Hence the areas of the triangles are almost the same.
The difference between the present computation and that
described above to estimate the accuracy
is that here on one boundary at $x=0$
we impose a fixed boundary condition
that prohibits the shape of the boundary from changing
from a wedge to a straight chord.
Thus, the resulting configurations have a wedge-shaped boundary at $x=0$, and
as $x$ becomes large, the shape approaches a string solution
(see Fig. \ref{dump.eps}).
If the difference in energy between the two branes of lengths $L=4$ and $8$
is sufficiently close to the energy of a confining string of
length $4=8-4$, we judge the configuration to be sufficiently close
to the string solution around $x\sim 4$, and we can use $L=4$ to compute the energies
of the baryon vertices.
\begin{table}[htb]
\caption{The numerically computed energies of the confining string configurations
with a fixed boundary at one end.
$T(n_{\rm mesh})$ is the numerically computed tension given in Table \ref{tension.tbl}.}
\label{dump.tbl}
\begin{center}
\begin{tabular}{c|rrrr}
\hline
\hline
$n_{\rm mesh}$ &
\multicolumn{1}{c}{$10$} &
\multicolumn{1}{c}{$20$} &
\multicolumn{1}{c}{$40$} &
\multicolumn{1}{c}{$80$} \\
\hline
$E_{L=8}$    & $87.2002$ & $87.75829$ & $87.88710$ & $87.91770$ \\
$E_{L=4}$    & $44.0562$ & $44.31901$ & $44.37818$ & $44.39194$ \\
\hline
$(E_{L=8}-E_{L=4})/(8-4)$ & $10.7860$ & $10.85982$ & $10.87723$ & $10.88144$ \\
Difference from $T(n_{\rm mesh})$ &  $+0.0002$ & $+0.00004$ & $+0.00001$ & $+0.00000$ \\
\hline
\end{tabular}
\end{center}
\end{table}
The result of this analysis is given in Table \ref{dump.tbl}.
There it is seen that indeed the configuration is
sufficiently close to a string solution before $x$ reaches $L=4$.

\begin{figure}[htb]
\centerline{\includegraphics[width=.9\linewidth]{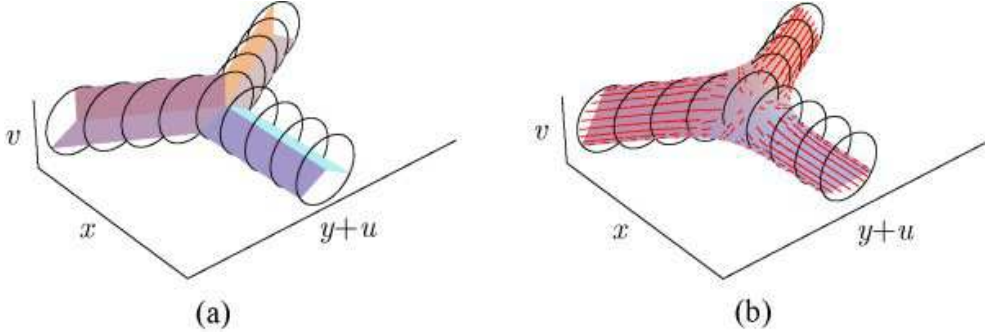}}
\caption{An initial configuration composed of three wedge configurations
and the corresponding final junction configuration obtained numerically.}
\label{junc.eps}
\end{figure}

Based on the above preliminary results,
we can finally compute the energies of the baryon vertices.
The initial configurations are combinations of three
wedges pasted like Witten's $3$-string vertices [see Fig. \ref{junc.eps} (a)].
The directions of three branches of initial configurations in the $(x,y)$-plane
are set as in Fig. \ref{balance.eps}, so that the tensions are balanced.
Now, instead of taking the $L\rightarrow\infty$ limit,
we simply use $L=4$, and each branch consists of $2n_{\rm mesh}^2$ triangles.

We give the results for the
KS background in Table \ref{juncks.tbl}.
Because vertices with more than three branches are unstable,
we only give the results for three string vertices.
\begin{table}[htb]
\caption{The energies of baryon vertices in the KS solution.
$E_{\rm junc}$ is the energy of an entire junction of branch length $L=4$.
$E_0$ represents the contribution of branches, computed
as $E_0=\sum_{i=1}^3LT(\theta_i^{\rm f})$.
The energy of a vertex is defined by $E_{\rm vertex}=E_{\rm junc}-E_0$.
The shaded numbers are the final results obtained by extrapolation.}
\label{juncks.tbl}
\begin{center}
\begin{tabular}{cc|rrrrr}
\hline
\hline
$(\theta_1^{\rm f},\theta_2^{\rm f},\theta_3^{\rm f})$ & $n_{\rm mesh}$ &
\multicolumn{1}{c}{$10$} &
\multicolumn{1}{c}{$20$} &
\multicolumn{1}{c}{$40$} &
\multicolumn{1}{c}{$80$} &
\multicolumn{1}{c}{$\infty$} \\
\hline
   $(\frac{4\pi}{12},\frac{4\pi}{12},\frac{4\pi}{12})$ & $E_{\rm junc}$           & $131.014$ & $131.8324$ & $132.0261$ & $132.0731$ & $132.0886$\hspace{0.27em} \\
$E_0=132.5661$ & $E_{\rm vertex}$ & $ -1.552$ & $ -0.7337$ &  $-0.5400$ &  $-0.5130$ &  \cbox{$-0.4775$} \\
\hline
   $(\frac{3\pi}{12},\frac{4\pi}{12},\frac{5\pi}{12})$ & $E_{\rm junc}$           & $128.126$ & $128.9356$ & $129.1274$ & $129.1740$ & $129.1894$\hspace{0.27em} \\
$E_0=129.6061$ & $E_{\rm vertex}$ &  $-1.480$ &  $-0.6705$ &  $-0.4787$ &  $-0.4321$ &  \cbox{$-0.4167$} \\
\hline
   $(\frac{2\pi}{12},\frac{5\pi}{12},\frac{5\pi}{12})$ & $E_{\rm junc}$           & $122.473$ & $123.3349$ & $123.5420$ & $123.5927$ & $123.6095$\hspace{0.27em} \\
$E_0=123.8908$ & $E_{\rm vertex}$ &  $-1.418$ &  $-0.5559$ &  $-0.3488$ &  $-0.2981$ &  \cbox{$-0.2813$} \\
\hline
\end{tabular}
\end{center}
\end{table}
Contrary to the intuitive expectation,
the signatures of the baryon vertex energies are negative.
Their absolute values
depend on the ratio of the flux angles,
and it seems that the energy becomes closer to zero as the flux angles
become increasingly asymmetric.
This behavior was expected because in the limit that one of the flux angles vanishes,
the junction becomes a confining string configuration, and the energy of the vertex goes to zero.

The results for the vertex energies in the MN background are more interesting.
As shown in Table \ref{junc.tbl},
the ratios of the vertex contributions to the
total energies are smaller than $10^{-5}$.
This strongly suggests that the energies of baryon vertices
exactly vanish for any values of the flux angles.
\begin{table}[htb]
\caption{The energies of baryon vertices in the MN solution.
$E_{\rm junc}$, $E_0$, and $E_{\rm vertex}$ are defined in the same way as those in Table \ref{juncks.tbl}.
The shaded numbers are the final results obtained by extrapolation.}
\label{junc.tbl}
\begin{center}
\begin{tabular}{cc|rrrrr}
\hline
\hline
$(\theta_1^{\rm f},\theta_2^{\rm f},\theta_3^{\rm f})$ & $n_{\rm mesh}$ &
\multicolumn{1}{c}{$10$} &
\multicolumn{1}{c}{$20$} &
\multicolumn{1}{c}{$40$} &
\multicolumn{1}{c}{$80$} &
\multicolumn{1}{c}{$\infty$} \\
\hline
   $(\frac{4\pi}{12},\frac{4\pi}{12},\frac{4\pi}{12})$ & $E_{\rm junc}$           & $129.584$ & $130.3545$ & $130.5356$ & $130.5794$ & $130.5939$\hspace{0.27em} \\
$E_0=130.5936$ & $E_{\rm vertex}$ &  $-1.010$ &  $-0.2391$ &  $-0.0580$ &  $-0.0142$ & \cbox{$+0.0003$} \\
\hline
   $(\frac{3\pi}{12},\frac{4\pi}{12},\frac{5\pi}{12})$ & $E_{\rm junc}$           & $126.616$ & $127.3866$ & $127.5686$ & $127.6127$ & $127.6273$\hspace{0.27em} \\
$E_0=127.6270$ & $E_{\rm vertex}$ &  $-1.011$ &  $-0.2404$ & $ -0.0584$ &  $-0.0143$ & \cbox{$+0.0003$} \\
\hline
   $(\frac{2\pi}{12},\frac{5\pi}{12},\frac{5\pi}{12})$ & $E_{\rm junc}$           & $121.116$ & $121.9670$ & $122.1716$ & $122.2217$ & $122.2383$\hspace{0.27em} \\
$E_0=122.2382$ & $E_{\rm vertex}$ &  $-1.122$ &  $-0.2712$ & $ -0.0666$ &  $-0.0165$ & \cbox{$+0.0001$} \\
\hline
\end{tabular}
\end{center}
\end{table}
In other words,
the energy of a junction is obtained by
summing up the contribution of each branch given as the product of
its tension and its length in the $(x,y)$-plane.
In Table \ref{batten.tbl}, we give another result
for a stable planar four-string vertex.
It too seems consistent with a vanishing vertex energy.
\begin{table}[htb]
\caption{The energy of a planar four-string vertex with $\theta_i^{\rm f}=\pi/4$.
$E_{\rm junc}$, $E_0$, and $E_{\rm vertex}$ are defined in the same way as those in Table \ref{juncks.tbl} and \ref{junc.tbl}.
The shaded number is the final result obtained by extrapolation.}
\label{batten.tbl}
\begin{center}
\begin{tabular}{cc|rrrrr}
\hline
\hline
$(\theta_1^{\rm f},\theta_2^{\rm f},\theta_3^{\rm f},\theta_4^{\rm f})$
 & $n_{\rm mesh}$ &
\multicolumn{1}{c}{$10$} &
\multicolumn{1}{c}{$20$} &
\multicolumn{1}{c}{$40$} &
\multicolumn{1}{c}{$80$} &
\multicolumn{1}{c}{$\infty$} \\
\hline
$(\frac{3\pi}{12},\frac{3\pi}{12},\frac{3\pi}{12},\frac{3\pi}{12})$ & $E_{\rm junc}$            & $141.329$ & $141.9459$ & $142.1079$ & 142.1537 & $142.1695$\hspace{0.27em} \\
$E_0=142.1723$ & $E_{\rm vertex}$ &  $-0.843$ & $ -0.2264$ &  $-0.0644$ & $-0.0186$ & \cbox{$-0.0028$}\\
\hline
\end{tabular}
\end{center}
\end{table}

\begin{figure}[htb]
\centerline{\includegraphics[width=.38\linewidth]{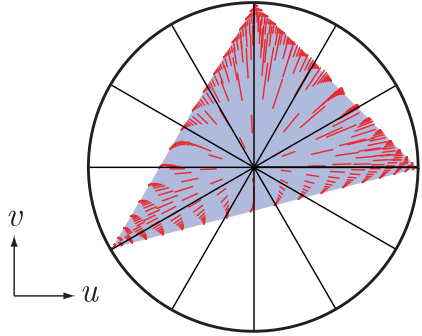}}
\caption{A three-string junction configuration
in the MN background projected on the $(u,v)$-plane.}
\label{triangle.eps}
\end{figure}

Before ending this section,
we confirm that junction configurations
in the MN background have the asymptotic forms suggested
in \S\ref{string.sec}, which are represented as triangles in the
$(u,v)$-plane.
The projection of a numerically generated junction in the $(u,v)$-plane is shown in Fig. \ref{triangle.eps},
and it indeed seems to be a triangle, as expected.

\section{Analytic proof}\label{analytic.sec}
As we saw in the last section,
the numerical results for junctions in the MN background
strongly suggest that baryon vertices do not
contribute to the energies of junctions.
This is also the case for supersymmetric junctions
consisting of $(p,q)$-strings.
\cite{schwarz,interaction,dasmuk,network,BPSpq,Mnetwork,matsuoku}
This fact for $(p,q)$-junctions is proven by
dualizing junctions to membrane configurations
in M-theory\cite{Mnetwork}.
From this viewpoint,
junctions are represented as
two-dimensional smooth surfaces embedded in
a four-dimensional space ${\bf R}^2\times{\bf T}^2$.
Let $(X,Y)$ and $(U,V)$ be
orthogonal coordinates in ${\bf R}^2$ and ${\bf T}^2$, respectively.
We combine them into two complex coordinates, $Z_1=X+iU$ and $Z_2=Y+iV$.
Then, a surface in this space is represented by
a function $Z_2=F(Z_1,\ol Z_1)$.
It is known that if we require a brane configuration to be supersymmetric,
the function $F$ must be holomorphic\cite{Mnetwork,matsuoku}.
This fact is very important in proving
the disappearance of the energies of
vertices\cite{Mnetwork}.

Brane configurations
in the MN background have a similar property.
As we have seen, they are described as two-dimensional smooth open surfaces
in the four-dimensional space ${\bf R}^2\times{\bf D}^2$
with coordinates $(x,y,u,v)$.
Let us define complex coordinates $z_1=x+iu$ and $z_2=y+iv$ in analogy to
the $(p,q)$-junction case.
A surface can be represented by an equation $z_2=f(z_1,\ol z_1)$.
Far from the vertex, each branch of the surface asymptotically approaches
a confining string solution,
which is represented by
\begin{equation}
y=a_ix+b_i,\quad
u=c_iv+d_i,
\label{factorize}
\end{equation}
where $i$ is the label of the branch.
When we rewrite these relations in terms of complex variables
$z_1$ and $z_2$, the function $f$ is holomorphic
if the coefficients $a_i$ and $c_i$ are the same.
We can always realize $a_i=c_i$ for one $i$ by applying the appropriate
rotation of the coordinates $u$ and $v$.
Furthermore,
due to the similarity of the triangles displayed in Fig. \ref{balance.eps} and Fig. \ref{baryonuv.eps},
the relation $a_i=c_i$ can be realized for all $i$ simultaneously.
Therefore, the function $f$ becomes holomorphic in the asymptotic part of
all the branches.
We call this property ``asymptotic holomorphy''.
In terms of real variables,
the holomorphy of the function $f$ is
represented by the Cauchy-Riemann relations:
\begin{equation}
\left(\frac{\partial v}{\partial u}\right)_x
=\left(\frac{\partial y}{\partial x}\right)_u,\quad
\left(\frac{\partial y}{\partial u}\right)_x
=-\left(\frac{\partial v}{\partial x}\right)_u.
\label{CR}
\end{equation}
In fact, the second equation in (\ref{CR}) is vacuous as a
condition for the asymptotic shapes of surfaces.
Because we are considering brane configurations which asymptotically approach
confining string solutions,
the asymptotic forms are always represented
by the two factorized equations in (\ref{factorize}).
Hence, here, the second equation in (\ref{CR}) reduces to
the trivial relation $0=0$.

\begin{figure}[htb]
\centerline{\includegraphics{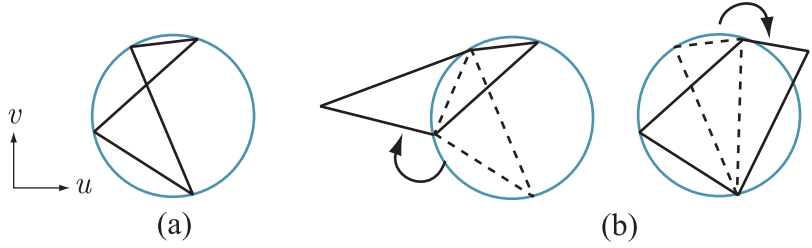}}
\caption{(a) A twisted square for $\theta_{\rm tot}^{\rm f}=2\pi$.
(b) Two ways of unfolding.}
\label{unfold.eps}
\end{figure}

This asymptotic holomorphy can be used to check
the stability of four-string vertices.
Under the assumption of the coincidence of the endpoints of chords,
the relation $\theta^{\rm f}=\theta^{\rm t}$ for the MN background implies that
$k$ chords representing branches of a $k$-string junction form a $k$-gon inscribed in the unit circle in
the $(u,v)$-plane.
For four-string junctions,
there are three cases distinguished by their values of $\theta_{\rm tot}^{\rm f}\equiv\sum_{i=1}^4\theta_i^{\rm f}$.
If $\theta_{\rm tot}^f=\pi$ or $3\pi$,
the squares defined in the $(u,v)$-plane in this way are
similar to the squares
composed of the four tension vectors of four external strings,
just as in the case of three-string junctions.
However, if $\theta^f_{\rm tot}=2\pi$,
the inscribed squares are twisted and cannot be similar to the
tension squares [see Fig. \ref{unfold.eps} (a)].
In this case, we obtain squares similar to the tension squares by unfolding
the twisted squares.
The two ways of unfolding correspond to the two different topologies (H) and (I),
which cannot be deformed smoothly into each other [see Fig. \ref{unfold.eps} (b)].
From these facts, we conclude that
a four-string vertex is stable only if it satisfies the
asymptotic holomorphy condition,
which is equivalent to the similarity of the tension squares and the
inscribed squares in the $(u,v)$-plane.

The asymptotic holomorphy relations are
not sufficient to compute the energies of baryon vertices.
For this purpose, we need to find some relations that hold everywhere on branes.
One possibility is to extend the asymptotic holomorphy
to the entire worldvolume as they are.
Unfortunately, however,
a numerical investigation of generated junction solutions demonstrates
that this cannot be done.

In the case of supersymmetric configurations,
there is a technique to determine relations holding everywhere
on the surface by using Killing spinors.
Indeed, various supersymmetric embeddings of D5-branes in the MN background
are obtained in Ref. \citen{flavoring} with this technique.
Unfortunately, this is not applicable to junction configurations,
because they break all the supersymmetries, as mentioned in \S\ref{string.sec}.

The absence of systematic methods compels us to resort to guesswork.
Numerically generated junction configurations give important information about
how to approach to this problem.
Let us consider Fig. \ref{triangle.eps}.
The radial pattern of the flux in it implies that $(D^u,D^v)\propto(u,v)$.
The proportionality factor can be determined by the Gauss's Law constraint (\ref{d1gauss}),
and we obtain the ansatz
\begin{equation}
D'_1=2\pi(udv-vdu),\quad
D'^a=\frac{2\pi}{\sqrt{\det g_{ab}}}u^a,
\label{Dansatz}
\end{equation}
where $u^a=(u,v)$ and we use the static coordinates $\sigma^a=u^a$.
Then, by substituting this ansatz into the energy (\ref{2dime}),
we obtain
\begin{equation}
E=2\pi\int d^2\sigma\sqrt{\det\wh g_{ab}},
\label{effectiveaction}
\end{equation}
with the effective induced metric $\wh g_{ab}$ defined by
\begin{equation}
\wh g_{ab}=\frac{1}{w}\delta_{ab}+wX^i_aX^i_b,
\label{effectiveg}
\end{equation}
where $X^i=(x,y)$ and $X^i_a=\partial X^i/\partial u^a$.
The expression (\ref{effectiveaction})
implies that the two-dimensional surfaces can be treated as
Nambu-Goto-type branes in the background with the effective metric
\begin{equation}
\wh{ds}^2=\frac{1}{w}(du^2+dv^2)+w(dx^2+dy^2).
\end{equation}
Thus, it is quite natural to replace $du$, $dv$, $dx$, and $dy$ in the Cauchy-Riemann relations (\ref{CR})
by $w^{-1/2}du$, $w^{-1/2}dv$, $w^{1/2}dx$, and $w^{1/2}dy$,
respectively.
As a result, we obtain
the following ``modified Cauchy-Riemann relations'',
which take the place of the holomorphy
for $(p,q)$-string junctions:
\begin{equation}
\left(\frac{\partial v}{\partial u}\right)_x
=\left(\frac{\partial y}{\partial x}\right)_u,\quad
w\left(\frac{\partial y}{\partial u}\right)_x
=-\frac{1}{w}\left(\frac{\partial v}{\partial x}\right)_u.
\label{modCR}
\end{equation}
Note that these relations do not contradict the asymptotic holomorphy relations
(\ref{CR})
but, rather, guarantee them.
Because the relation between $(x,u)$ and $(y,v)$ is factorized as
(\ref{factorize}) in asymptotic regions,
the second relation in (\ref{modCR})
becomes trivial as the second relation in (\ref{CR}).

We can show that the equations of motion are
automatically satisfied if the shape of a brane
satisfies the modified Cauchy-Riemann relations
(\ref{modCR}) and the flux density on it
is given by the ansatz (\ref{Dansatz}).
The equation of motion for $X^i$
obtained from the energy (\ref{effectiveaction})
is
\begin{equation}
\partial_a\left(\sqrt{\det\wh g_{ab}}w\wh g^{ab}X^i_b\right)=0.
\label{Xeom}
\end{equation}
In order to obtain the equation of motion for the gauge field,
we should use the action (\ref{2dime}),
to which the ansatz (\ref{Dansatz}) has not been applied.
With variations in the form
$\delta D'^a=(1/\sqrt{g})\epsilon^{ab}\partial_b\phi$,
which do not violate the constraint (\ref{d1gauss}),
we obtain the Maxwell's equation for the electric field strength $E_a$,
\begin{equation}
\epsilon^{ab}\partial_a E_b=0,
\label{maxwellE}
\end{equation}
where the relation between $E_a$ and $D^a$ is
\begin{equation}
E_a=\frac{g_{ab}D^a}{\sqrt{(2\pi w)^2+g_{ab}D^aD^b}}.
\label{eledef}
\end{equation}
This can be rewritten using the
ansatz (\ref{Dansatz})
as
\begin{equation}
E_a
=\frac{g_{ab}u^b}{\sqrt{\det\wh g_{ab}}}
=\frac{\wh g_{ab}u^b}{w\sqrt{\det\wh g_{ab}}}.
\label{Eans}
\end{equation}

To show that equations of motion (\ref{Xeom}) and (\ref{maxwellE}) hold,
we first rewrite the relations (\ref{modCR}) in terms of
the functions $x(u,v)$ and $y(u,v)$.
Then, using the chain rule for partial derivatives,
we obtain
\begin{equation}
x_u+y_v=0,\quad
x_uy_v-y_ux_v=\frac{1}{w^2},
\label{holomorphyxyuv}
\end{equation}
or equivalently,
\begin{equation}
\tr X_a^i=0,\quad
\det X_a^i=\frac{1}{w^2}.
\end{equation}
(Here and hereafter, when subscripts are used to represent derivatives,
the independent variables are always $u$ and $v$.)
Using these relations,
the effective metric (\ref{effectiveg}) becomes
\begin{equation}
\wh g_{ab}=w(X_a^iX_b^i+\delta_{ab}\det X_a^i)
=w(x_v-y_u)\left(\begin{array}{cc}
-y_u & -y_v \\
x_u & x_v
\end{array}\right),
\label{gbym}
\end{equation}
and its determinant is
\begin{equation}
\det\wh g_{ab}=(x_v-y_u)^2.
\label{detghat}
\end{equation}
With the help of (\ref{gbym}) and (\ref{detghat}),
we can easily show that
the equations of motion (\ref{Xeom}) and (\ref{maxwellE})
are indeed satisfied.

Although we have shown that the modified Cauchy-Riemann relations (\ref{modCR})
and the ansatz (\ref{Dansatz})
guarantee that brane configurations
satisfy the equations of motion (\ref{Xeom}) and (\ref{maxwellE}),
the converse is not true,
as there exist solutions of equations of motion
which do not satisfy the modified Cauchy-Riemann relations
and the $D$-field ansatz.
An example of such a configuration is the deformed baryon configuration
given in Ref. \citen{deform}.
Such configurations, however, do not approach confining string
solutions asymptotically,
and their existence does not invalidate the arguments
given in this section.

\begin{figure}[htb]
\centerline{\includegraphics[width=.9\linewidth]{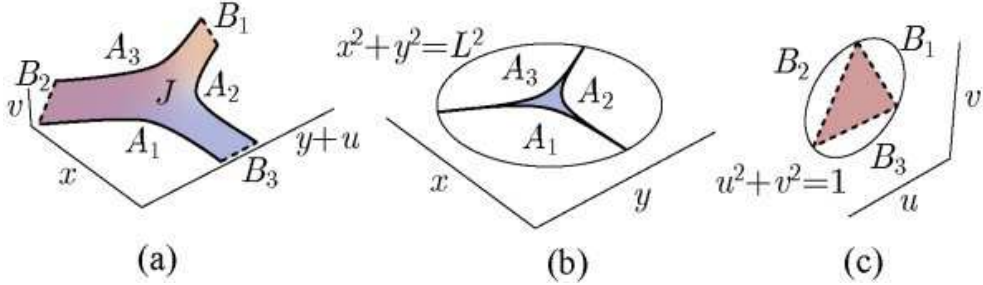}}
\caption{(a) $J$ and its boundary. (b) The projection on the $(x,y)$-plane. (c) The projection on the $(u,v)$-plane.}
\label{energy.eps}
\end{figure}

Now, we are ready to prove the vanishing of
the energies of baryon vertices analytically.
From (\ref{effectiveaction}) and (\ref{detghat}),
the energy $E_{\rm junc}(L)$ defined above (\ref{ebdef}) is given by
\begin{equation}
E_{\rm junc}(L)=2\pi\int_J du dv|x_v-y_u|
        =2\pi\left|\oint_{\partial J}(xdu+ydv)\right|,
\label{analytice}
\end{equation}
where $J$ is the region in a junction worldvolume determined by
the relation $x^2+y^2\leq L^2$.
Before applying Stokes' theorem to (\ref{analytice}), the absolute
value should be moved out of the integral.
This is possible, because as shown by (\ref{holomorphyxyuv}),
$x_v-y_u$ never vanishes on the worldvolume.
The boundary $\partial J$ consists of two kinds of
boundaries, which we call A-boundaries (the solid lines in Fig. \ref{energy.eps})
and B-boundaries (the dashed lines in Fig. \ref{energy.eps}).
The A-boundaries are the edges of the surface on the boundary $u^2+v^2=1$ of the target space.
The B-boundaries arise due to the cut off $x^2+y^2=L$.
The A-boundaries are mapped to apices of the triangle when
they are projected on the $(u,v)$-plane,
and the coordinates $u$ and $v$ are constants on them.
Therefore,
the A-boundaries do not contribute to the contour integral
in (\ref{analytice}).
Thus, we need only consider the integral on the B-boundaries.
If $L$ is sufficiently large, each B-boundary is mapped to a point
by the projection in the $(x,y)$-plane.
Let $(x_i,y_i)$ be the point corresponding to the $i$-th B-boundary, $B_i$.
These are on the circle $x^2+y^2=L^2$,
and we can write
\begin{equation}
(x_i,y_i)=L(\alpha_i,\beta_i),\quad(\alpha_i^2+\beta_i^2=1)
\label{xyab}
\end{equation}
where $(\alpha_i,\beta_i)$ is a unit vector representing
the direction of the $i$-th branch in the $(x,y)$-plane.
If we project the B-boundaries to the $(u,v)$-plane, they are mapped to sides of the triangle
when $L$ is sufficiently large.
Due to the similarity between this triangle and the tension triangle,
which is guaranteed by the modified Cauchy-Riemann relations (\ref{modCR}),
the integration measure $(du_i,dv_i)$ on $B_i$
can be represented as
\begin{equation}
(du_i,dv_i)=\pm(\alpha_i,\beta_i)dl_i,
\label{uvab}
\end{equation}
where the signature depends on the orientation of the contour,
and $dl_i$ is an infinitesimal length on the $i$-th side
of the triangle in the $(u,v)$-plane.
Combining (\ref{xyab}) and (\ref{uvab}), we obtain the following expression
for the energy of a junction:
\begin{equation}
E_{\rm junc}=2\pi\sum_i L\int_{B_i}dl_i=2\pi\sum_i L\times 2\sin\theta_i^{\rm f}=\sum_i LT(\theta^{\rm f}_i).
\label{analyticejunc}
\end{equation}
Here, we have used the fact that the integral $\int_{B_i} dl_i$ gives the length of
the $i$-th side of the triangle $2\sin\theta_i^{\rm f}$
and the tension formula (\ref{ttheta}) for the MN background.
The result (\ref{analyticejunc}) implies that
the energy of a junction is given by the sum of
the energies of the branches and that there is no vertex contribution.

Now we can answer the question raised at the end of \S\ref{IR.sec}.
Although the expression (\ref{EFT2}) seemingly
depends on $g_{\rm str}$, which has no counterpart in the boundary field
theory of the MN solution,
$\wt E_{\rm vertex}$ on the right-hand side
is actually zero, and the energies of
the baryon vertices in fact are independent of $g_{\rm str}$.

\section{Conclusions}\label{conc.sec}
In this paper, we have investigated string junctions in ${\cal N}=1$
supersymmetric gauge theories
in the context of the gauge/gravity correspondence.
We have used the Maldacena-N\'u\~nez and the Klebanov-Strassler solutions
as gravity duals of the confining gauge theories.

In \S\ref{string.sec} we studied the balance of the tensions
for three-string junctions and planar four-string junctions.
We found that four-string vertices are stable only in the case that
$\theta_{\rm tot}^{\rm f}=\pi$ or $3\pi$ in the MN solution.
Planar four-string vertices in the KS solution are always unstable.
It may be interesting to generalize this consideration to non-planar junctions.

In \S\ref{numerical.sec}, we reported the results of numerical computations of
the energies of baryon vertices, and we found that in the KS case the energies are negative, while
in the MN case they almost vanish.
The vanishing of the energies strongly suggests that the brane configurations in the MN background
possess some
analytic structure
which guarantees this vanishing.
Indeed, we discovered relations similar to the Cauchy-Riemann relations
for holomorphic surfaces.
With the help of these relations, we analytically proved the disappearance of
the energies of baryon vertices.

We should emphasize that the analysis given in this paper is classical.
As mentioned in \S\ref{string.sec}, the existence of confining strings
breaks all the supersymmetries.
Thus, there is no mechanism to control quantum corrections,
and our results are justified only in the limit of large $N$ and
large $Ng_{\rm str}$, in which the background curvature is small
compared to the string scale and the Plank scale.

In order to investigate realistic baryons in non-supersymmetric
QCD, we have to use different gravity backgrounds.
For example, we can use an AdS Schwarzchild black hole\cite{thermal,HP}.
Because the IR geometry of this solution has structure similar
to the KS and MN solutions, we can study junction configurations in it
in a manner similar to that used in this paper.

We have treated only brane configurations embedded in the IR geometry
in this paper.
They can be used for highly-excited baryons, in which
the endpoints of strings are separated from each other.
To analyze ground-state baryons, we should consider different
kinds of brane configurations that consist of a baryon vertex
and only one external string going up along the $r$ direction to
flavor branes. The external string in this case represents coincident quarks.
It is important to determine whether such branes are stable, and if so, to investigate their energies and excitations.

There are many interesting problems
associated with brane constructions of hadrons
in addition to those mentioned above.
We hope to return to these issues in the near future.

\section*{Acknowledgements}
I would like to thank M.~Bando and A.~Sugamoto for motivating me to do this work.
I also thank F.~Koyama, H.~Ooguri, S.~Sugimoto and M.~Tachibana for valuable discussions.
This work is supported in part by
a Grant-in-Aid for the Encouragement of Young Scientists
(\#15740140) from the Japan Ministry of Education, Culture, Sports,
Science and Technology,
and by the Rikkyo University Special Fund for Research.

\appendix
\section{}
\subsection{Treatment of flux on triangulated surfaces}
The purpose of this section is
to explain how to descretize branes with flux flowing on them and
how to vary variables while maintaining the Gauss's Law constraint.

\begin{figure}[htb]
\centerline{\includegraphics{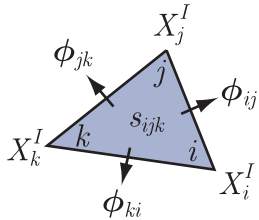}}
\caption{The triangle $ijk$. $X^I_i$ denotes the coordinate of
the vertex $i$. $\phi_{ij}$ represents the flux flowing across the link $ij$.
The area of the triangle $ijk$ is denoted by $s_{ijk}$.}
\label{trigabc.eps}
\end{figure}

To compute the energy of a D3-brane, which is represented in (\ref{2dime}) as
an integral over a two-dimensional surface,
we first triangulate the surface.
We label sites by $i,j,k,\ldots$ and each oriented link by
two labels representing its two ends.
The functions $X^I(\sigma^a)$ describing the shape of the surface
are replaced by the variables $X^I_i=(x_i,y_i,u_i,v_i,w_i)$
on the sites,
with the constraints $u_i^2+v_i^2+w_i^2=1$ and $w_i\geq0$.
To represent the flux density, we use link variables
\begin{equation}
\phi_{ij}\equiv\int_i^jD'_1,
\end{equation}
where the integral is carried out along the link $ij$.
This represents the amount of flux flowing across the link $ij$.
We define the orientation in such a way that if the arrow from site $i$ to $j$
is upward, $\phi_{ij}$ represents the flux passing the link $ij$ from
left to right.
By definition we have
\begin{equation}
\phi_{ij}=-\phi_{ji}.
\end{equation}
The area of a triangle $ijk$ is denoted by $s_{ijk}$
(see Fig. \ref{trigabc.eps}).
This area is a real positive number.
We also define $s'_{ijk}$ as the area of the triangle $ijk$ projected on the $(u,v)$-plane.
This take either a positive or negative value, depending on
the orientation of the triangle.
$s_{ijk}$ and $s_{ijk}'$ are easily represented as functions of
the variables $X_i^I$, $X_j^I$ and $X_k^I$.

Given the variables $X_i^I$ and $\phi_{ij}$, the energy of the D3-brane is
obtained as the discretized version of the energy (\ref{2dime}),
\begin{equation}
E=\sum_{\rm triangles} s_{ijk}\sqrt{(2\pi w_{ijk})^2+\sum_I(D'^I_{ijk})^2},
\end{equation}
where $w_{ijk}$ is the $w$-component of the center of mass $X_{ijk}^I=(X_i^I+X_j^I+X_k^I)/3$
 of the triangle $ijk$,
and $D'^I_{ijk}$ is the push-forward of the electric flux density to the five-dimensional space $(x,y,u,v,w)$,
which is given by
\begin{equation}
D'^I_{ijk}=\frac{1}{2s_{ijk}}\{\phi_{jk}(X^I_i-X^I_{ijk})+\phi_{ki}(X^I_j-X^I_{ijk})+\phi_{ij}(X^I_k-X^I_{ijk})\}.
\label{trigD}
\end{equation}

A discretized version of the Gauss's Law constraint is
\begin{equation}
\phi_{ij}+\phi_{jk}+\phi_{ki}=2\pi\rho s'_{ijk}.
\label{gausslaw}
\end{equation}
In order to find a configuration that minimizes the energy,
we should vary the variables $X^I_i$ and $\phi_{ij}$
in such manner that does not violate this Gauss's Law constraint.

\begin{figure}[htb]
\centerline{\includegraphics{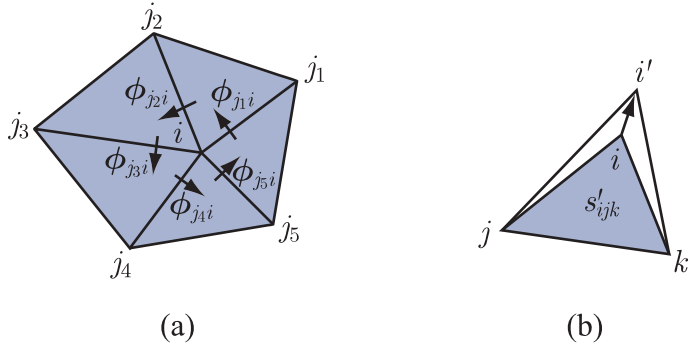}}
\caption{
(a) A variation changing the rotation of the electric flux density
around $i$
while maintaining the Gauss's Law constraint.
(b) If a site $i$ moves to $i'$,
the projected areas of the triangles that possesses this site as a corner
change.}
\label{pentamovei.eps}
\end{figure}

There are two kinds of variations.
Variations of $\phi_{ij}$ with fixed $X_i^I$ are
generated by the following variation for each site $i$:
\begin{equation}
\phi_{ji}\rightarrow\phi_{ji}+c,\quad j\in{\rm Adj}(i).
\label{vari1}
\end{equation}
Here, ${\rm Adj}(i)$ represents the set of all the sites
adjoining the site $i$.
This variation changes the rotation of the flux density around the
site $i$ [see Fig. \ref{pentamovei.eps} (a)].
The other kind of variations are those that change $X_i^I$.
Even when we vary the variables $X^I_i$,
we should take account of the Gauss's Law constraint (\ref{gausslaw}),
because
variations of $X^I_i$ change charges in
triangles.
If the position of a site $i$ moves from $X^I_i$ to
$X^I_{i'}=X^I_i+\delta X^I_i$,
the projected area $s'_{ijk}$ of the triangle $ijk$
is changed by the amount $s'_{jii'}-s'_{kii'}$
[see Fig. \ref{pentamovei.eps} (b)].
To maintain the Gauss's Law constraint (\ref{gausslaw}),
we must change the
flux variables simultaneously according to the relation
\begin{equation}
\phi_{ki'}=\phi_{ki}-2\pi\rho s'_{kii'},\quad
k\in{\rm Adj}(i).
\label{vari2}
\end{equation}
Any continuous deformation can be generated by
the two kinds of variations (\ref{vari1}) and (\ref{vari2}).

\end{document}